\documentclass[prX,showpacs,nofootinbib]{revtex4}

\usepackage{amssymb}    
\usepackage{amsbsy}
\usepackage{amsmath}
\usepackage{amsfonts}
\usepackage{mathrsfs}
\usepackage{latexsym}
\usepackage[english]{babel}
\usepackage{url}
\newcommand{\feyn}[1]{#1\kern-0.45em/}
\begin{document}
% ----------------------------------------------------------------

\title{All orders renormalizability of a Lorentz and $CPT$ violating quantum electrodynamics}
%\title{On the gauge anomalies of a Lorentz and $CPT$ violating quantum electrodynamics}
%\title{On the anomalies of a Lorentz-violating quantum electrodynamics}
%\title{All orders renormalization of a Lorentz-violating quantum electrodynamics}

\author{O.M. Del Cima}
\email{oswaldo.delcima@ufv.br}
\affiliation{Universidade Federal de Vi\c cosa (UFV),\\
Departamento de F\'\i sica - Campus Universit\'ario,\\
Avenida Peter Henry Rolfs s/n - 36570-000 -
Vi\c cosa - MG - Brazil.}

\author{J.M. Fonseca}
\email{jakson.fonseca@ufv.br}
\affiliation{Universidade Federal de Vi\c cosa (UFV),\\
Departamento de F\'\i sica - Campus Universit\'ario,\\
Avenida Peter Henry Rolfs s/n - 36570-000 -
Vi\c cosa - MG - Brazil.}
\affiliation{Universidade Federal de Vi\c{c}osa (UFV),\\
Campus de Rio Parana\'{i}ba,\\ 
Rodovia MG-230 Km 7 - 38810-000 - Rio Parana\'{i}ba - MG - Brazil.}

\author{D.H.T. Franco}
\email{daniel.franco@ufv.br}
\affiliation{Universidade Federal de Vi\c cosa (UFV),\\
Departamento de F\'\i sica - Campus Universit\'ario,\\
Avenida Peter Henry Rolfs s/n - 36570-000 -
Vi\c cosa - MG - Brazil.}

\author{A.H. Gomes}
\email{andre.herkenhoff@ufv.br}
\affiliation{Universidade Federal de Vi\c cosa (UFV),\\
Departamento de F\'\i sica - Campus Universit\'ario,\\
Avenida Peter Henry Rolfs s/n - 36570-000 -
Vi\c cosa - MG - Brazil.}

\author{O. Piguet}
\email{opiguet@pq.cnpq.br} 
\affiliation{Universidade Federal do Esp\'\i rito Santo (UFES),\\
CCE, Departamento de F\'\i sica,\\
Campus Universit\'ario de Goiabeiras - 29060-900 - Vit\'oria - ES - Brazil.}
\affiliation{Universidade Federal de Vi\c cosa (UFV),\\
Departamento de F\'\i sica - Campus Universit\'ario,\\
Avenida Peter Henry Rolfs s/n - 36570-000 -
Vi\c cosa - MG - Brazil.}
%\author[ufv]{O.M. Del Cima\corref{cor1}}
%\ead{oswaldo.delcima@ufv.br}

%\author[ufv,ufv-par]{J.M. Fonseca}
%\ead{jakson.fonseca@ufv.br}

%\author[ufv]{D.H.T. Franco}
%\ead{daniel.franco@ufv.br}

%\author[ufv]{A.H. Gomes}
%\ead{andre.herkenhoff@ufv.br}

%\author[ufes,ufv]{O. Piguet}
%\ead{opiguet@yahoo.com} 

%\cortext[cor1]{Corresponding author.}
%\address[ufv]{Universidade Federal de Vi\c{c}osa (UFV),
%Departamento de F\'{i}sica, Campus Universit\'{a}rio,
%Avenida Peter Henry Rolfs s/n - 36570-000 - Vi\c{c}osa - MG - Brazil.}
%\address[ufv-par]{Universidade Federal de Vi\c{c}osa (UFV), Campus de Rio Parana\'{i}ba, 
%Rodovia MG-230 Km 7 -38810-000 - Rio Parana\'{i}ba - MG - Brazil.}
%\address[ufes]{Universidade Federal do Esp\'{i}rito Santo (UFES),
%Departamento de F\'{i}sica,
%Campus Universit\'{a}rio de Goiabeiras - 29060-900 - Vit\'{o}ria - ES - Brazil.}

\date{\today}

%===================================================================
\begin{abstract}

Renormalizability of the (minimal) single-fermion QED extension is investigated at all orders of perturbation theory in the framework of algebraic renormalization, a regularization-independent method. Relative to the standard QED, new structures that could lead to gauge anomalies are identified. Nevertheless, even if the anomaly coefficients fail to vanish in the general case, they shall be absent provided we require invariance of the action under $C$ and/or $PT$ transformations. Stability is also verified in this case, hence full renormalizability is attained.

\end{abstract}

%\begin{keyword}
%Lorentz symmetry breaking \sep renormalization \sep gauge anomaly
%\end{keyword}

%Renormalization in field theory, 11.10.Gh, 11.10.Hi
%Gauge field theory, 11.15.-q
%Perturbation theory applied to gauge field theories, 11.15.Bt
%Symmetry in theory of fields and particles, 11.30.-j

\maketitle
%===================================================================

%%%%%%%%%%%%%%%%%%%%%%%%%%%%%%%%%%%%%%%%%%%%%%
\section{Introduction}
%%%%%%%%%%%%%%%%%%%%%%%%%%%%%%%%%%%%%%%%%%%%%%

Quantum gravity effects, coming from Planck scale, may appear as small violations of fundamental laws in the limit of low energies and in the last two decades a great deal of effort has been put in the possibility of breaking both Lorentz and $CPT$ symmetries: detection of such effects could help pave the way towards a consistent quantum theory of gravity \cite{annual-review}.

An example of a systematic approach which has been intensively studied is the Standard Model Extension (SME) \cite{sme}, a Lorentz and $CPT$ violating\footnote{An important result states that violations of $CPT$ symmetry in a (local) quantum field theory \cite{streater-wightman} must be accompanied with the loss of Lorentz invariance, although the converse may not be true \cite{greenberg}.} extension of the Standard Model. This extension is an effective low-energy limit theory comprising all the possible deviations from the Standard Model arising from high-energy fundamental theories with Lorentz covariant dynamics in which spontaneous Lorentz violation may occur. The minimal extension respects SU(3)$\times$SU(2)$\times$U(1) gauge symmetry and is power-counting renormalizable -- eventually, further requirements like causality, unitarity, etc., may be imposed, yielding more restricted models. High precision tests in various sectors of the SME have already bounded many of the breaking coefficients \cite{data-table}, but with no evidence for the violation of Lorentz symmetry so far.

In this work, we consider the issue of renormalizability of the (minimal) single-fermion QED extension. At one-loop order, a proof of multiplicative renormalizability was given in \cite{kost-1-loop}. Here, renormalizability will be studied at \textit{all orders} of perturbation theory in the algebraic approach \cite{piguet}, a regularization-independent method. Despite being an effective model, this kind of study is important because it inevitably looks at the unitarity of the theory and, if we are to expect the high energy behavior to be unitary, any non-unitarity appearing in the low energy effective regime would signal a limit of the domain of validity of this approximation. We find new anomaly structures besides the usual Adler-Bardeen-Bell-Jackiw one. Although the latter remains under control thanks to the Adler-Bardeen non-renormalization theorem \cite{adler-bardeen-theorem}, the remaining anomalies are potentially dangerous since no analogous theorem is known, which would guarantee their absence from an eventual vanishing of their one-loop order coefficients. To any extent, we show that restricted models which are $C$ and/or $PT$ invariant are definitely free of anomalies. Stability -- meaning that all renormalization ambiguities are equivalent to a redefinition of the parameters of the theory -- is checked in this anomaly free case, completing the proof of its renormalizability.

%%%%%%%%%%%%%%%%%%%%%%%%%%%%%%%%%%%%%%%%%%%%%%
\section{Extended QED: Classical approach}
%%%%%%%%%%%%%%%%%%%%%%%%%%%%%%%%%%%%%%%%%%%%%%

\subsection{The model}%%%%%%%%%%%%%%%%%%%%%%%%

In the tree approximation, the action we work with is given by \cite{sme}:
\begin{equation}\label{action}
\mathcal{S} = \mathcal{S}_{\mathrm{QEDex}} + \mathcal{S}_{\mathrm{GF}} + \mathcal{S}_{\mathrm{IR}},
\end{equation}
where $\mathcal{S}_{\mathrm{QEDex}}$ is the extension, with Lorentz and $CPT$ symmetry breakings, of the action of QED for electrons 
and photons,
\begin{eqnarray}\label{qedex}
\mathcal{S}_{\mathrm{QEDex}} = \int d^{\,4}x\, \left[	i\, \overline{\psi}\, \Gamma^\mu D_\mu\, \psi - \overline{\psi}\, M\, \psi - \frac{1}{4}F^{\mu\nu} F_{\mu\nu} - \frac{1}{4}(k_F)_{\alpha\beta\kappa\lambda} F^{\alpha\beta}F^{\kappa\lambda} + (k_{AF})_\mu A_\nu \widetilde{F}^{\mu\nu}	\right],
\end{eqnarray}
which, besides the usual QED  terms, includes Lorentz breaking terms whose coefficients have the form of constant background fields\footnote{A linear operator $A_\mu$ coupled with a background field $(k_A)_\mu$ could also be present but would introduce linear instabilities in the potential and is therefore assumed to vanish at tree-level (radiative corrections to this term are not expected to be present -- see first reference of \cite{sme}).}. 
We have used the definitions:
\begin{equation}
\Gamma^\mu \equiv \gamma^{\,\mu} + c^{\,\mu\nu}\gamma_\nu + d^{\mu\nu}\gamma_5\gamma_\nu + e^{\,\mu} + 
if^\mu\gamma_5 + \frac{1}{2}\,g^{\alpha\beta\mu}\sigma_{\alpha\beta},
\end{equation}
\begin{equation}
M \equiv m + im_5\gamma_5 + a^{\,\mu}\gamma_\mu + b^{\,\mu}\gamma_5\gamma_\mu + \frac{1}{2}\,H^{\mu\nu}\sigma_{\mu\nu}.
\end{equation}
We note that constant tensor fields of even number of indexes respect $CPT$ symmetry and the ones with odd number do not, both violating (active) Lorentz invariance once they give rise to preferential directions in spacetime, breaking its isotropy. The coefficient $(k_F)_{\alpha\beta\kappa\lambda}$ and those appearing in $\Gamma^\mu$ are dimensionless, while $(k_{AF})_\mu$ and the ones in $M$ have dimensions of mass, all of them being real because of the reality of the action. The gauge-fixing action is of the Stueckelberg type \cite{ECGS},
\begin{equation}
	\mathcal{S}_{\mathrm{GF}} = \int d^{\,4}x\, \left[	-\frac{1}{2\alpha}\left(\partial_\mu A^\mu\right)^2	\right],
\end{equation}
and the infrared regulator action, introduced in order to avoid infrared (IR) singularities, 
\begin{equation}
	\mathcal{S}_{\mathrm{IR}} = \int d^{\,4}x\, \left(	\frac{\mu^2}{2}A_\mu A^\mu	\right),
\end{equation}
is a mass term for the photon field -- gauge Ward identities are not spoiled by the photon mass, a peculiarity of the Abelian case \cite{low-schroer}.

On experimental grounds, already breaking coefficients are known to be very small -- possibly leading to Planck-suppressed effects -- in any Earth-based frames of reference or other inertial frames with low velocity relative to Earth \cite{data-table}. To avoid spurious enlargement of these parameters, we restrict ourselves to these frames. Also, Lorentz violation effects may be of the same order of magnitude as higher loop corrections and, for consistence of the approach, since we perform analyses at all orders in perturbation theory, we consider contributions of arbitrary order in these coefficients.

\subsection{Classical symmetries}%%%%%%%%%%%%%%%%%%%%%%%%

Not only $CPT$ is lost, but none of the discrete operations $C$, $P$ or $T$ is a symmetry of the model (see Table \ref{table:table1}, where the coefficients represent the associated field operators) and since Lorentz is also broken, invariance under U(1) gauge transformations is the only exact symmetry of the extended QED action. Variations under this transformation are functionally implemented by the gauge Ward operator,
\begin{equation}\label{gaugewardop}
W_{\rm g} = \int d^{\,4}x \,\Lambda(x) w_{\rm g}(x),
\end{equation}
with,
\begin{equation}
w_{\rm g}(x) = -\partial^{\,\mu} \frac{\delta}{\delta A^\mu} + ie\left(	\frac{\overleftarrow{\delta}}{\delta\psi}\,\psi - \overline{\psi}\,\frac{\overrightarrow{\delta}}{\delta\overline{\psi}}	\right),
\end{equation}
where $\Lambda(x)$ is the infinitesimal gauge transformation parameter. The action (\ref{action}) therefore changes under infinitesimal 
gauge transformations as:
\begin{equation}\label{gaugeward-integ}
W_{\rm g} \,\mathcal{S} = - \int d^{\,4}x \,\Lambda(x) \left( \frac{\Box+\alpha\mu^2}{\alpha} \right)\partial_\mu A^\mu.
\end{equation}
Note that this breaking of the gauge invariance is linear in $A^\mu$ and therefore will not be renormalized, remaining a classical breaking \cite{piguet,low-schroer}. The local form of (\ref{gaugeward-integ}),
\begin{equation}\label{gaugeward}
w_{\rm g}(x)\,\mathcal{S} = - \left( \frac{\Box+\alpha\mu^2}{\alpha} \right)\partial_\mu A^\mu
\end{equation}
is the classical gauge Ward identity.

\begin{table}[ht]
\centering
\caption{Discrete-symmetry properties of the field operators.}
\begin{tabular}{l cc cc cc cc cc cc cc}
\hline
\hline
				&& $\textrm{C}$ && $\textrm{P}$ && $\textrm{T}$ && $\textrm{CP}$ && $\textrm{CT}$ && $\textrm{PT}$ && $\textrm{CPT}$ \\
	\hline
$c_{00},(k_F)_{0j0k},$		 	 &&		&&	   &&     &&	 &&		&&	   &&		\\
\,\,$c_{jk},(k_F)_{jklm},$		 && $+$ && $+$ && $+$ && $+$ && $+$ && $+$ && $+$	\\
%	\hline
$c_{0j},c_{j0},(k_F)_{0jkl}$	 && $+$ && $-$ && $-$ && $-$ && $-$ && $+$ && $+$	\\
%	\hline
$b_j,g_{j0l},g_{jk0},(k_{AF})_j$ && $+$ && $+$ && $-$ && $+$ && $-$ && $-$ && $-$	\\
%	\hline
$b_0,g_{j00},g_{jkl},(k_{AF})_0$ && $+$ && $-$ && $+$ && $-$ && $+$ && $-$ && $-$	\\
%	\hline
$a_0,e_0,f_j$	 				 && $-$ && $+$ && $+$ && $-$ && $-$ && $+$ && $-$	\\
%	\hline
$a_j, e_j, f_0$					 && $-$ && $-$ && $-$ && $+$ && $+$ && $+$ && $-$	\\
%	\hline
$H_{jk},d_{0j},d_{j0}$			 && $-$ && $+$ && $-$ && $-$ && $+$ && $-$ && $+$	\\
%	\hline
$H_{0j},d_{00},d_{jk}$			 && $-$ && $-$ && $+$ && $+$ && $-$ && $-$ && $+$	\\
\hline
\hline
\end{tabular}
\label{table:table1}
\end{table}

It should be stressed that in the model we are considering here -- contrary to the case presented in \cite{qed-missing-analysis}, where Lorentz symmetry is broken only by a soft breaking term -- Lorentz symmetry is fully broken by the most general terms respecting gauge invariance, hermiticity, therefore, Lorentz Ward identities are meaningless.   

When dealing with the issue of quantization of the model, one asks whether or not its symmetries survive after this process. If they do not, one says that there are anomalies. In the following, we study this issue for the U(1) gauge symmetry within the algebraic method of renormalization \cite{piguet}. The anomaly issue, in the case of a gauge symmetry, is physically crucial because of its well known link with the unitarity of the corresponding quantum theory.

%%%%%%%%%%%%%%%%%%%%%%%%%%%%%%
\section{Quantization}
%%%%%%%%%%%%%%%%%%%%%%%%%%%%%%

The algebraic renormalization approach is based on two fundamental steps: (i) the study of the Wess-Zumino consistency condition in order to verify that the quantization of the model does not destroy any classical symmetry, \textit{i.e.}, no anomaly is present; and (ii) the analysis of the stability of the action, guaranteeing that it is the most general power-counting renormalizable action obeying the symmetries of the model, ensuring that all counterterms will be properly reabsorbed by a redefinition of the parameters of the starting action. Verification of both proves the existence of a renormalized theory fulfilling the Ward identity (\ref{gaugeward}) together with suitable normalization conditions \cite{piguet,piguet-rouet}  fixing the free parameters.

In order to define a perturbative expansion one has to split the classical action (\ref{action}) in a free and an interacting part. Since the Lorentz breaking terms are supposed to be small on physical grounds, it appears reasonable to consider all of them, including the ones which are quadratic in the quantum fields, as interactions. We shall also limit the degree in these external fields to a fixed finite number, 
for the same physical reason. The practical consequence of the latter assumption is that it avoids the occurrence of an infinite number of
Feynman graphs having the same number of loops. We can therefore define as usually the expansion order as the number of loops, equivalent to the power in the Planck constant $\hbar$.

An other most important consequence of the Lorentz invariance of the free action, hence of the free propagators, is that it is an essential assumption in the proofs of the Quantum Action Principle (QAP) available in the literature \cite{LamClarke,Low-MaiBreit,BrenDutsch}, which are given for Poincar\'e invariant theories. The presence of Lorentz breaking interaction vertices does not spoil these proofs, thus we can apply the QAP to the present case. We have of course to suppose we are using a subtraction scheme of the UV singularities, such as BPHZ, dimensional regularization or Epstein-Glaser renormalization, for which the QAP has been proved \cite{LamClarke,Low-MaiBreit,BrenDutsch}.

\subsection{Wess-Zumino consistency condition}%%%%%%%%%%%%%%%%%%%%%%%%

The loop or $\hbar$ expansion of the vertex functional $\Gamma$, the generating functional of the 1-particle irreducible graphs,
\begin{equation}\label{hbar-exp}
	\Gamma = \sum_{n \, \geq \, 0} \hbar^n \, \Gamma^{(n)} = \mathcal{S} + \mathcal{O}(\hbar),
\end{equation}
is such that it coincides with the classical action (\ref{action}) in the classical limit: $\Gamma^{(0)} = \mathcal{S}$.

The vertex functional $\Gamma$ of the quantum theory we want to define should obey the Ward identity (\ref{gaugeward}) -- with $\Gamma$ instead of $\mathcal{S}$. If not, it is called anomalous. The construction begins by considering a vertex functional $\Gamma$ expanded as in (\ref{hbar-exp}). Its variation under the local Ward operator (\ref{gaugewardop}) yields, according to the QAP:
\begin{eqnarray}\label{qap}
w_{\rm g}(x)\, \Gamma +  \left( \frac{\Box+\alpha\mu^2}{\alpha} \right)\partial_\mu A^\mu(x) = \Delta_{\rm g}(x) \cdot \Gamma = \Delta_{\rm g}(x) + \mathcal{O}(\hbar\Delta_{\rm g}),
\end{eqnarray}
where $\Delta_{\rm g}(x)$ is a local polynomial of the fields with UV dimension bounded by $d_{\mathrm{UV}} \leq 4$.

Applying to $\Gamma$ the commutation rule satisfied by the local gauge Ward operator,
\begin{equation}
	[w_{\rm g}(x),w_{\rm g}(y)]=0,
\end{equation}
we obtain the Wess-Zumino consistency condition,
\begin{equation}\label{wesszumino}
w_{\rm g}(x)\Delta_{\rm g}(y)-w_{\rm g}(y)\Delta_{\rm g}(x)=0.
\end{equation}
A ``trivial'' solution for $\Delta_{\rm g}(x)$ reads:
\begin{equation}\label{trivial}
\Delta_{\rm g} (x) = w_{\rm g} (x) \hat{\Delta}_{\rm g},
\end{equation}
where $\hat{\Delta}_{\rm g}$ is an integrated (gauge noninvariant) field polynomial. With $\Delta_{\rm g}(x)$ written as (\ref{trivial}), the counterterms can be chosen recursively in such a way that the Ward identity is satisfied at any order and the theory is free of anomalies. Otherwise, if there is any non-trivial term composing $\Delta_{\rm g}(x)$, i.e. it can not be written as (\ref{trivial}), it represents a \textit{potential} gauge anomaly.

The search for possible anomalies now reduces to the task of listing all polynomials composing $\Delta_{\rm g}(x)$, with the restriction of $d_{\textrm{UV}} \leq 4$. These are of the form, symbolically:
\begin{eqnarray}\label{basis}
A^4, \partial A^3, \partial^{\,2} A^2, \partial^{\,3} A, A^3, \partial A^2, \partial^{\,2} A, A^2, \partial A, A, \partial\, (\overline{\psi}\,\psi), A\, \overline{\psi}\,\psi, \overline{\psi}\,\psi,
\end{eqnarray}
(considering all possible contractions among indexes, possibly using $\gamma$-matrices, Levi-Civita symbols $\varepsilon$ and/or the background fields) -- and verifying if they (or combinations among them) can or not all be written in the form (\ref{trivial}). Concomitantly, some polynomials may also be excluded from $\Delta_{\rm g}(x)$ due other specific demands (\textit{e.g.}, discrete symmetries).

Investigation of all terms in (\ref{basis}) reveals that polynomials involving derivatives can be cast in the form of (\ref{trivial}) as long as there are no antisymmetric contractions with $\partial_\alpha A_\beta$; also, $A^4$, $A^3$, $\varepsilon A\,\partial A$, $A^2$, $A$ and $A\,\overline{\psi}\,\psi$ do not satisfy the Wess-Zumino condition (\ref{wesszumino}) and are immediately excluded from the list (\ref{basis}). Therefore, $\Delta_{\rm g}(x)$ is given by:
\begin{eqnarray}\label{candidate.anomalies}
\Delta_{\rm g}(x) = \Big\{ \overline{\psi}\,\psi, \varepsilon \partial A \partial A, \varepsilon\partial A \Big\}
\end{eqnarray}
(where the $\varepsilon$'s are reminders of antisymmetric contractions with $\partial A$). If no extra conditions are imposed to remove these polynomials, they represent \textit{potential} anomalies.

In the standard QED, discrete symmetries play a \textit{crucial role}, ruling out all polynomials that can not be of the trivial form (\ref{trivial})\footnote{For example, the gauge Ward operator (\ref{gaugewardop}) is odd under charge conjugation, implying in this case, from (\ref{qap}), that $\Delta_{\rm g}(x)$ must also be. This removes all $C$ invariant terms from $\Delta_{\rm g}(x)$. For more details, see Section 5 of \cite{piguet-rouet}.}. Here we have no discrete symmetries and the terms of (\ref{candidate.anomalies}) can not be written as (\ref{trivial}), being candidates for anomalies, so the (quantum) Ward identity reads, writing (\ref{candidate.anomalies}) more explicitly:
\begin{eqnarray}\label{ward-anomalies}
w_{\rm g}(x)\, \Gamma  & = & - \left( \frac{\Box+\alpha'\mu'^{\,2}}{\alpha'} \right)\partial_\mu A^\mu + \lambda^{(1)}\overline{\psi}\psi +\, i\lambda^{(2)}\overline{\psi}\gamma_5\psi +\, \lambda^{(3)}_\mu\overline{\psi}\gamma^\mu\psi +\, \lambda^{(4)}_\mu\overline{\psi}\gamma^\mu\gamma_5\psi +\, \lambda^{(5)}_{\mu\nu}\overline{\psi}\,\sigma^{\mu\nu}\psi	\nonumber	\\
&& +\, \lambda^{(6)}_{\mu\nu\rho\sigma} F^{\mu\nu}F^{\rho\sigma} +\, \lambda^{(7)}_{\mu\nu}F^{\mu\nu} +\, \mathcal{O}\left(\hbar\lambda\right),
\end{eqnarray}
where the coefficients $\alpha'$ and $\mu'$ are possible renormalizations of $\alpha$ and $\mu$, and the anomaly coefficients $\lambda^{(i)}$ are functions of the parameters of the theory defined in the action (\ref{action}). Note that the first term in the right-hand side 
satisfies the consistency condition (\ref{wesszumino}). The Adler-Bardeen-Bell-Jackiw (ABBJ) anomaly is a special case of the sixth anomaly term, with $\lambda^{(6)}_{\mu\nu\rho\sigma}$ proportional to the Levi-Civita tensor $\varepsilon_{\mu\nu\rho\sigma}$.

Discrete symmetries -- \textit{e.g.} $C$ -- usually eliminate anomalous terms. Here, in the absence of any discrete symmetry, we are left with  all these \textit{new kinds of possible anomalies}, whose coefficients $\lambda^{(i)}$ may be calculated via, for instance, explicit evaluation of the associated Feynman diagrams\footnote{We know $W_{\rm g}$, therefore by applying functional derivatives in (\ref{ward-anomalies}) (and setting the fields to zero) we are able to find the diagrams that contribute to each anomaly coefficient.}: determination of $\lambda^{(1)}$ to $\lambda^{(5)}$ hinges on the  computation of $\delta^3 \Gamma / (\delta\overline{\psi}\delta\psi\delta A^\mu)$ (``vertex correction" diagram); $\lambda^{(6)}$ depends on $\delta^3 \Gamma / (\delta A^\mu\delta A^\nu\delta A^\rho)$ (``triangle" diagram); and $\lambda^{(7)}$ is determined by $\delta^2 \Gamma / (\delta A^\mu\delta A^\nu)$ (``photon self-energy" diagram). Some of these coefficients could be vanishing, as is the case for the ABBJ anomaly in the Standard Model due to special cancellations. However, in the present case, there is no non-renormalization theorem available (analogous to the Adler-Bardeen theorem \cite{adler-bardeen-theorem}), which could control the anomalies out of one-loop order calculations. Hence, we do not have a serious motivation to  undertake the task of computing the latters.

\subsection{Discrete symmetries $C$ and/or $PT$}%%%%%%%%%%%%%%%%%%%%%%%%

Now we turn our attention to the possibility of removing these potential anomalies by requiring (separately) $C$ and/or $PT$ symmetry of the action (\ref{action}). For definiteness, henceforth we choose $PT$ symmetry -- the other two cases go analogously. As can be checked in Table I, imposing this invariance requires the absence of the coefficients $b^{\mu}$, $d^{\mu\nu}$, $g^{\alpha\beta\mu}$, $H^{\mu\nu}$, and $(k_{AF})^\mu$. % \cite{field-redefinition}. 
In more details: if these background fields are not present, the action recovers $PT$ invariance and, along with the fact the gauge Ward operator (\ref{gaugewardop}) is $PT$-odd, from (\ref{qap}) it is clear that $\Delta_{\rm g}(x)$ must also be $PT$-odd. This requirement removes all candidate anomalies (\ref{ward-anomalies}) from $\Delta_{\rm g}(x)$, % \cite{footnote-discr-rem}, 
and therefore the $PT$ invariant model is guaranteed to be anomaly-free (all coefficients at the r.h.s. of (\ref{ward-anomalies}) now vanish).

%In passing we also remark that besides $C$ and/or $PT$ symmetry no other discrete invariance combining $C$, $P$ and/or $T$ could be imposed on the %action (\ref{action}) so that it is free of any potential anomaly and still Lorentz violating: coordinate invariance while keeping (active) Lorentz %violation allows imposition of only $C$ and/or $PT$ symmetries and $CPT$ symmetry as a whole is just too broad to rule out all terms from %(\ref{candidate.anomalies}). 

As a side note, it is important to deal with care the requirement of discrete symmetries even if the gauge Ward identity (\ref{ward-anomalies}) turns out to be truly anomalous: when considering the full QED extension, with all fermion families, the possibility of anomaly cancellations may emerge (see first reference of \cite{sme}) and this may avoid the necessity of requiring the vanishing of the individual fermion contributions to the coefficients. As already mentioned, this would also require the existence of a non-renormalization theorem.
 
\subsection{Stability}%%%%%%%%%%%%%%%%%%%%%%%%

In the present context, once gauge invariance is proven to hold to all orders --- e.g., thanks to one of the discrete symmetries already mentioned ---, \textquotedblleft{stability}\textquotedblright means that radiative corrections can all be reabsorbed by a redefinition of the parameters of the theory. We keep on with the $PT$ symmetry as above. It is well-known that the stability of the quantum perturbative theory is guaranteed if the classical theory (the classical action together with the classical Ward identity) itself is stable \cite{piguet} under small perturbations of dimension less or equal to 4. We therefore perform such a perturbation, $\epsilon\,\widetilde{S}$ ($\epsilon \ll 1$) on the ($PT$ invariant) action, $\mathcal{S} \rightarrow \mathcal{S} + \epsilon\,\widetilde{\mathcal{S}},$ and by requiring this perturbed action to satisfy the classical gauge Ward identity (\ref{gaugeward}):
\begin{eqnarray}
w_{\rm g}(x) (\mathcal{S} \,+\, \epsilon\,\widetilde{\mathcal{S}}) = w_{\rm g}(x)\, \mathcal{S} \,+\, \epsilon\,w_{\rm g}(x)\, \widetilde{\mathcal{S}} \equiv - \left( \frac{\Box+\alpha\mu^2}{\alpha} \right)\partial_\mu A^\mu,
\end{eqnarray}
we conclude that all possible counterterms must be gauge invariant, $W_{\rm g}\, \widetilde{\mathcal{S}} \equiv 0$, and $PT$ even (so we guarantee the absence of anomalies). This selects precisely the $PT$ invariant terms $\mathscr{P}_i$ of the classical action (\ref{action}):
\begin{alignat*}{2}
	&	\mathscr{P}_1 = i\overline{\psi}\,\gamma^{\mu} D_\mu \psi,
	%\quad
	&&	\mathscr{P}_2 = i\overline{\psi} \,c^{\,\mu\nu} \gamma_\nu D_\mu\psi,
	\\
	\displaybreak[0]
	&	\mathscr{P}_3 = i\overline{\psi} \,(k_F)^{\mu \lambda \nu}_{\,\,\,\,\,\,\,\,\,\,\lambda} \gamma_\nu D_\mu\psi,
	\qquad	
	&&	\mathscr{P}_4 = i\overline{\psi}\,e^{\,\mu} D_\mu\psi,
	\\
	\displaybreak[0]
	&	\mathscr{P}_5 = \overline{\psi}\,f^{\mu} \gamma_5\, D_\mu\psi,
	%\quad	
	&&	\mathscr{P}_6 = \overline{\psi}\,\psi,
	\\
	\displaybreak[0]
	&	\mathscr{P}_7 = \overline{\psi} \,a^{\,\mu} \gamma_\mu\,\psi,
	%\quad
	&&	\mathscr{P}_8 = \overline{\psi} \,e^{\,\mu} \gamma_\mu\,\psi,
	\\
	\displaybreak[0]
	&	\mathscr{P}_9 = F^{\mu\nu} F_{\mu\nu},
	%\quad
	&&	\mathscr{P}_{10} = (k_F)_{\alpha\beta\kappa\lambda} F^{\alpha\beta} F^{\kappa\lambda},
	\\
	\displaybreak[0]
	&	\mathscr{P}_{11} = c_{\alpha\beta\kappa\lambda} F^{\alpha\beta} F^{\kappa\lambda},
	%\quad
	&&	\mathscr{P}_{12} = i \overline{\psi}\,\gamma_5\,\psi,
\end{alignat*}
where $c_{\alpha\beta\kappa\lambda} \equiv \eta_{\alpha\lambda}(c_{\beta\kappa}+c_{\kappa\beta}) -\eta_{\alpha\kappa}(c_{\beta\lambda}+c_{\lambda\beta}) - \eta_{\beta\lambda}(c_{\alpha\kappa}+c_{\kappa\alpha}) + \eta_{\beta\kappa}(c_{\alpha\lambda}+c_{\lambda\alpha})$, which have the same symmetries of $(k_F)_{\alpha\beta\kappa\lambda}$.

Finally, the most general integrated local function $\widetilde{\mathcal{S}}$ which is gauge and $PT$ invariant is given by:
\begin{equation}
	\widetilde{\mathcal{S}} = \int d^{\,4}x \sum_{i=1}^{12} a_i \mathscr{P}_i (x),
\end{equation}
where $a_1, \dots, a_{12}$ represent renormalizations of the coefficients of the gauge and PT invariant action. Note that these coefficients remain arbitrary. They may be fixed by normalization conditions at the classical order, and by induction, order by order in perturbation theory. This ends the proof of the renormalizability of this theory. A quite similar proof holds for the $C$ invariant theory.

We thus saw that full renormalizability demands $C$ and/or $PT$ invariance. Choosing $C$ symmetry rules out $a_\mu$, $d_{\mu\nu}$, $e_{\mu}$, $f_\mu$, and $H_{\mu\nu}$ from the original action (\ref{action}). If $PT$ symmetry is choosen, $b_{\mu}$, $d_{\mu\nu}$, $g_{\alpha\beta\mu}$, $H_{\mu\nu}$, and $(k_{AF})_\mu$ need to be absent. With both $C$ and $PT$ invariance, there only remain $c_{\mu\nu}$ and $(k_F)_{\alpha\beta\kappa\lambda}$. %Comparing these cases we find that the $C$ and $PT$ odd coefficients $d_{\mu\nu}$ and $H_{\mu\nu}$ are expected to vanish in the single-fermion QED extension.

%%%%%%%%%%%%%%%%%%%%%%%%%%%%%%
\section{Conclusions and Prospects}
%%%%%%%%%%%%%%%%%%%%%%%%%%%%%%

In summary, by means of the algebraic method, we identified all possible \textit{candidate} anomalies in the (minimal) single-fermion QED extension --- these would come from the ``vertex correction", ``triangle" and ``photon self-energy" diagrams. Explicit evaluation of the anomaly coefficients are left for future works but in practice may demand new non-renormalization theorems. The theory was proved to be renormalizable at \textit{all orders} in perturbation theory under the hypothesis of the existence of a discrete symmetry, namely $C$ and/or $PT$.

%%%%%%%%%%%%%%%%%%%%%%%%%%%%%%
\section*{Acknowledgements}
%%%%%%%%%%%%%%%%%%%%%%%%%%%%%%
The authors gratefully acknowledge V.A. Kosteleck\'{y} for many useful comments. O.M.D.C. dedicates this work to his children, Vittoria and Enzo, and to his mother, Victoria. This work was supported by CAPES and CNPq.

%%%%%%%%%%%%%%%%%%%%%%%%%%%%%%
%\section*{References}

%\bibliographystyle{model1-num-names}

%%%%%%%%%%%%%%%%%%%%%%%%%%%%%%
%%%%%%%%%%%%%%%%%%%%%%%%%%%%%%

% ----------------------------------------------------------------
\end{document}